\input harvmac

\Title{\vbox{\baselineskip12pt
\hbox{astro-ph/0303422}}}
{\vbox{\centerline{Dark Matter as Dark Energy}}}

\baselineskip=12pt \centerline {Ramzi R. Khuri\footnote{$^*$}
{e-mail: khuri@gursey.baruch.cuny.edu.}}
\medskip
\centerline{\sl Department of Natural Sciences, Baruch College, CUNY} 
\centerline{\sl 17 Lexington Avenue, New York, NY 10010}
\medskip
\centerline{\sl Graduate School and University Center, CUNY}
\centerline{\sl 365 5th Avenue, New York, NY 10036}

\bigskip
\centerline{\bf Abstract}
\medskip
\baselineskip = 20pt

Velocity-dependent interactions in a fundamental-string dominated universe lead quite
naturally, with reasonable assumptions on initial conditions, to an accelerating expanding universe 
without assuming the existence of a cosmological constant.
This result also holds generically for a universe dominated by moving extremal black holes, owing
to a repulsive velocity-dependent force. This interaction, however, does not preclude structure
formation. Here we discuss a toy model including both ordinary and extremal matter, in which the latter 
accounts for dark matter while simultaneously acting as effective dark energy. Eternal acceleration is
once more seen to arise in this case.

\Date{March 2003}

\def\r{\rho}

\def\({\left (}
\def\){\right )}
\def\[{\left [}
\def\]{\right ]}

\lref\cosmo{R. R. Khuri, Phys. Lett. {\bf B548} (2002) 9, hep-th/0209049.}

\lref\fund{R. R. Khuri, Phys. Lett. {\bf B520} (2001) 353, hep-th/0109041.}

\lref\shell{R. R. Khuri and A. Pokotilov, Nucl. Phys. {\bf B633} (2002)295, hep-th/0201194;
Phys. Lett. {\bf B535} 1 (2002) hep-th/0202158.}

\lref\ferr{R. C. Ferrell and D. M. Eardley, Phys. Rev. Lett. {\bf 59} (1987) 1617.}

\lref\bps{M. K. Prasad and C. M. Sommerfield, Phys. Rev. Lett. {\bf 35}
(1975) 760; E. B. Bogomol'nyi, Sov. J. Nucl. Phys. {\bf 24} (1976) 449.}

\lref\hawk{S. W. Hawking, Phys. Rev. {\bf D15} (1977) 2738.}

In \refs{\fund,\shell}, it was shown that the velocity-dependent 
forces between parallel fundamental strings or $p$-branes lead to an 
accelerating, expanding universe in the transverse space. In this model, the repulsive
force effectively
acts as a positive cosmological constant (dark energy) without explicitly
introducing a cosmological constant into the field equations. 
More recently \cosmo, the results of \refs{\fund,\shell}\ were
generalized to include a more general class of velocity-dependent Lagrangians.
Such Lagrangians are the generic interaction for multi-soliton solutions, whose static
interaction vanishes due to the cancellation of gauge and gravitational forces
(the zero force condition \bps).
In particular, these findings are not peculiar to string theory, and hold for,
say, the velocity-dependent interactions of moving extremal Reissner-Nordstr\" om
black holes \ferr. An accelerating expanding universe arises whenever
the velocity-dependent Lagrangian has the form
\eqn\lagone{L = {m v^2 \over 2} \left( 1 + f(r)v^2 \right),}
as in the case of maximally supersymmetric strings or branes up to
$O(v^4)$ in \refs{\fund,\shell}, or of the form
\eqn\lagtwo{L = {m v^2 \over 2} \left( 1 + g(r) \right),}
as in the case of extremal RN black holes \ferr, whenever $f(r)$ and $g(r)$ are 
monotonically decreasing functions of
the separation $r$. 
It was further shown in \cosmo\ that, while this type of velocity-dependent force is always 
repulsive, thus leading to an accelerating growth in the size of the model universe, this 
interaction also allows for cosmic structure formation, or clumping, in which the
velocity-dependent force essentially acts as a critical damping.

In this letter we discuss a toy model in which a matter-dominated universe consists of
ordinary matter, interacting only gravitationally, and ``extremal" matter, interacting 
gravitationally with the ordinary matter and via velocity-dependent interactions with other extremal matter.
In this model, the extremal matter represents dark matter, which, owing to the acceleration 
resulting from its repulsive velocity-dependent forces, also effectively accounts for 
dark energy. Since this matter is in the form of extremal black holes or similar
types of objects, the dark nature can be understood both from the black hole properties
of such bodies and from the fact that extremal black holes, having zero Hawking temperature
\hawk, also do not radiate.

Throughout our discussion we base our arguments on a highly simplified mean-field
approximation. From the results of \cosmo, we assume that structure formation is possible
due to the critical damping nature of the repulsive force. Suppose further that the matter
consists of $N$ clumps each consisting of mass $M$ of dark matter and mass $m$ of ordinary
matter. Assume $m < M$, and to leading order in $m/M$, the ordinary matter, being 
dominated by the dark matter, moves in clumps centered around the center
of mass of the clumps of dark matter. In a mean-field approximation, we take both types
of matter to be distributed as identical coincentric spheres moving apart in an expanding
universe. 
Let the mean separation between clumps be $R(t)$ and the mean size of each clump
to be $r(t)$. Then $r(t)$ also represents the mean distance of ordinary matter to 
the center of each clump. 

In order for the ordinary
matter to continue moving within each clump with the dark matter, the net attractive
force from all other clumps must be balanced by the net attractive force within each
clump. This implies
\eqn\ordbal{{G(M+m)\over r^2} \simeq {G(N-1) (M+m)\over R^2}.}
For large $N$, this requires 
\eqn\scales {r(t) \simeq {R(t)\over \sqrt{N}}.}
Similar considerations would apply to substructure within the clumps, thereby constraining
scale sizes in this manner.

For the overall motion of each clump, accelerated expansion occur
when the repulsive velocity-dependent dark matter-dark matter force overcomes the
attractive gravitational force between dark matter and ordinary matter.

Before addressing the total force on dark matter in this picture, consider first the case of a universe 
dominated by 
the dark matter. In \fund, a toy model consisting of only fundamental strings, $p$-branes or $D$-branes
was shown to lead naturally to an accelerating universe in the transverse space. Assuming
the constituents to be $N$ pointlike particles (say $D0$-branes) each of mass $M$, 
the mean-field model of \fund\ can be integrated explicitly for $D=4$ to yield \shell
\eqn\mf{\left({\r +3 \over \r +1}\right)\ln \left( \sqrt{R\over a} + \sqrt{{R\over a}+1}\right)
+ \sqrt{R\over a} \sqrt{{R\over a}+1} = \sqrt{\left(\r+2\right)^3\over \r\left(\r+1\right)^2}
\left({t-t_0\over k_1}\right) c,}
where $\r = E/M$ is the constant interaction energy per mass, $a=\r k_1/(\r +2)$
and 
\eqn\kay{k_1 = {G  N M\over 2\pi c^2}}
is the interaction scale of the velocity-dependent interaction.
The $\rho$-dependent factors are henceforth taken to be $O(1)$, which is manifestly
the case for all values of $\rho$.

In this model, $R(t)$ represents the mean position of
a single particle (or clump) with respect to the center of mass of the entire universe,
accounted for in the mean-field approximation as being concentrated at the center.
$R(t)$ therefore represents the mean size of the universe in this toy model, which consists only
of dark matter. Note that all factors of $c$ and $G$ have been restored in \mf\ as compared to the
same equation derived in \shell.

For small separation, $R << k_1$, or post-inflation cosmological times early compared to present, 
\eqn\quadratic{R(t) \sim {c^2 t^2 \over k_{1}},} 
implying a constant positive acceleration of
\eqn\mfearlyacc{\ddot R \sim {4\pi c^4 \over G M N}.}
For very large $R$, $R >> k_1$ (or late time), $R \simeq \sqrt{\r(\r +2)/(\r +1)^2} c t$,
so that the very late time acceleration eventually tends to zero, but always remains positive. 
It is straightforward to show from \mf\ that the subleading latetime accelerating has the
dependence
\eqn\mflateacc{\ddot R \sim {GNM\over R^2}.}

Now suppose we combine the extremal matter with the velocity dependent interaction
with ordinary matter, which interacts only gravitationally. 
The net acceleration on each clump of extremal matter is then the result of
adding the above repulsive acceleration between different extremal matter clumps
and the gravitational attraction of the ordinary matter, also situated in the
same clumps. The result for early times is
\eqn\main{\ddot R \sim -{GmN\over R^2} + {4\pi c^4 \over G M N}.}
It follows from \main\ that once an initially expanding universe
surpasses the size
\eqn\rzero{ R_0 \sim {GN\sqrt{mM}\over \sqrt{4\pi} c^2},}
roughly the geometric mean between the Schwarzschild radii of the ordinary and
dark matter, the repulsive force
begins to dominate the attractive force and the universe enters into a phase
of accelerated expansion.
Eventually, however, the repulsive force weakens with greater separation.
However, from \mflateacc, and for $M > m$, the repulsive acceleration 
$GNM/R^2$ still dominates
the gravitational attraction  $GNm/R^2$ and hence leads to eternal acceleration.

Of course this kind of toy model is highly simplified and need not to be taken
too seriously at present. However, it does show the essential feature of such
models, namely that the presence of repulsive
velocity-dependent forces, such as exist generically between extremal
black holes or similar classes of objects, can easily account for both dark
matter and an effective cosmological constant without the explicit introduction 
of this constant in the field equations. This would have the benefit of avoiding the
theoretical and conceptual problems associated with the cosmological constant and
de Sitter space. While large extremal black holes are astrophysically unrealistic,
the possibility that similar extremal matter (such as fundamental strings or related
objects) has the same interactive behaviour certainly makes such models worth further
exploration. The next step in this endeavour is then to consider 
the many-body problem for the case of the above two species of matter (ordinary
and extremal matter) . The possibility of a numerical
simulation along these lines is currently being investigated.

{\bf Acknowledgements:} 
I would like to thank Gia Dvali for suggesting the idea of considering
dark matter in this context. I would also like to thank Andriy Pokotilov 
and Xinle Yang for helpful discussions. 
Research supported by PSC-CUNY Grant \# 64535 00 33.

\listrefs

\end